\documentclass[aps,pra,twocolumn,superscriptaddress,a4paper,longbibliography]{revtex4-2}
\usepackage{graphicx}
\usepackage{natbib}
\usepackage{hyperref}
\usepackage{amsmath}
\usepackage{mathtools}
\usepackage{braket}
\usepackage{lipsum}
\usepackage{color}

\begin{document}


\title{Fano Resonance in Excitation Spectroscopy and\\ Cooling of an Optically Trapped Single Atom}

\author{Chang Hoong Chow}
\affiliation{Center for Quantum Technologies, 3 Science Drive 2, Singapore 117543}
\author{Boon Long Ng}
\affiliation{Center for Quantum Technologies, 3 Science Drive 2, Singapore 117543}
\author{Vindhiya Prakash}
\affiliation{Center for Quantum Technologies, 3 Science Drive 2, Singapore 117543}
\author{Christian Kurtsiefer}
\affiliation{Center for Quantum Technologies, 3 Science Drive 2, Singapore 117543}
\affiliation{Department of Physics, National University of Singapore, 2 Science Drive 3, Singapore 117542}
\email[]{christian.kurtsiefer@gmail.com}
\date{\today}

\begin{abstract}

Electromagnetically induced transparency (EIT) can be used to cool an atom in a harmonic potential close to the ground state by addressing several vibrational modes simultaneously.
Previous experimental efforts focus on trapped ions and neutral atoms in a standing wave trap.
In this work, we demonstrate EIT cooling of an optically trapped single neutral atom, where the trap frequencies are an order of magnitude smaller than in an ion trap and a standing wave trap.
We resolve the Fano resonance feature in fluorescence excitation spectra and the corresponding cooling profile in temperature measurements.
A final temperature of around 6\,$\mu$K is achieved with EIT cooling, a factor of two lower than the previous value obtained using polarization gradient cooling.

\end{abstract}


\maketitle
\section{Introduction}
Single neutral atoms in optical dipole traps form a potential basis for quantum information processing applications, including quantum simulation~\cite{Ebadi2021, Scholl2021}, computation~\cite{Bluvstein2022, Graham2022}, and communication~\cite{weinfurter2020,Thomas2022}.
Ideally, the atom can be prepared in an arbitrary quantum state and can be made to exchange quantum information coherently with a tightly focused optical mode.
A prerequisite for an efficient coupling between a photon and an atom is minimizing the atomic position uncertainty, which requires the atom to be sufficiently cooled~\cite{wilson2017}.
Furthermore, cooling of the atom can extend the coherence time of the qubit state ~\cite{Weiss2015,Chow2021,Graham2022} and allow for the manifestation of quantum mechanical properties of the atomic motion~\cite{Kaufman2014,Brown2023}.

Raman sideband cooling techniques~\cite{Monroe1995,Kaufman2012,Thompson2013} can be employed to cool atoms to the motional ground state of the trapping potential.
However, this method requires an iteration of the cooling process over several laser settings to address individual vibrational modes.
Alternatively, cooling by electromagnetically induced transparency (EIT) is a simpler approach that can also help achieve ground state cooling.

\begin{figure}
    \includegraphics[width=0.45\textwidth]{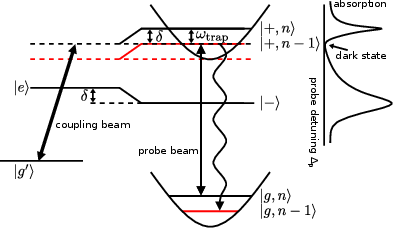}
    \caption{Left: EIT cooling transition in a three-level $\Lambda$ system.
    A strong coupling beam forms new eigenstates $\ket{+}$ and $\ket{-}$ from
    the bare atomic states $\ket{g'}$ and $\ket{e}$. Here 
    $n$ denotes the vibrational quantum number for atomic motion in a harmonic trap with a frequency of $\omega_\text{trap}$.
    By choosing a suitable intensity for the coupling beam, the scattering spectrum can be engineered such that the transition $\ket{g,n}\rightarrow\ket{+,n-1}$ is enhanced to achieve cooling. 
    Right: Spectral profile of the dressed states. 
    Scattering of a weak probe beam that couples an atom prepared in ground state $\ket{g}$ to the dressed states reveals two peaks corresponding to each of the dressed states and an asymmetric-Fano profile due to the dark state.
    \label{fig:dressedpicture}}
\end{figure}

EIT cooling relies on suppression of diffusion when a three-level atom is transferred to a superposition of the ground states that is decoupled from the excited state (dark state).
On probing the excitation spectrum of a $\Lambda$ system with a strong field (coupler) and a weaker probe, the dark state is revealed via a reduction in fluorescence when the probe and coupler are equally detuned from the excited state. 
This dip, in combination with the fluorescence peak from the dressed state, results in an asymmetric Fano profile~\cite{Cohen-Tannoudji_1992}. 
When the motional spread of the atomic wavepacket in an external conservative potential is taken into account, the dark state becomes sensitive to the atomic position. 
Particularly, cooling occurs when the dark state is decoupled from the excited state at the carrier frequency but is coupled to the bright (dressed) state at the red sideband~\cite{Morigi2003}. 
For this, the system simply needs to be engineered such that frequency difference between the dark state and the bright dressed state matches the vibrational mode spacing of the potential (see Fig.~\ref{fig:dressedpicture}).

The Fano profile was first observed in the fluorescence spectroscopy of a single Barium ion~\cite{spectroJanik1985,spectroStalgies1998}, and a
cooling technique exploiting this effect was proposed fifteen years later~\cite{Keitel2000}.
Since then, this EIT cooling method has been implemented in platforms such as trapped ions~\cite{Blatt2000,christian_longion_2016,monroe_longion_2020}, neutral atoms confined in standing wave traps~\cite{Meschede2014}, and quantum gas microscopy setups~\cite{Haller2015}.

In this work, we investigate free-space EIT cooling of a single neutral $^{87}$Rb atom in a mK deep far-off-resonant optical dipole trap (FORT), where the trap frequencies are typically around tens of kHz, one to two orders of magnitude smaller than in typical standing wave traps and ion traps.
A three-level $\Lambda$ system is realized using the magnetic sublevels in the hyperfine manifolds of the ground and excited states.
We first resolve the Fano profile via excitation spectroscopy, and then implement a cooling scheme on altering the configuration and detunings. 
We also explore the parameter space to identify detunings and
intensities that minimize the temperature.

\section{Theoretical Overview}
Theoretical descriptions of the Fano spectrum, and cooling by EIT have
been extensively reported earlier ~\cite{Cohen-Tannoudji_1992,
  spectroStalgies1998,Keitel2000, Morigi2003}.
Here we summarize the results and extend some of the outcomes to describe our measurements. 
Consider a $\Lambda$ system formed by two ground states $\ket{g}$ and $\ket{g'}$  as well as an excited state $\ket{e}$ that can decay to both ground states with a total decay rate $\Gamma$. 
A weak (strong) probe (coupling) field of frequency $\omega_p$ ($\omega_c$) couples $\ket{g}$ ($\ket{g'}$) to $\ket{e}$ with a Rabi frequency $\Omega_p$ ($\Omega_c$) and a detuning $\Delta_p = \omega_p - \omega_{eg}$ ($\Delta_c = \omega_c - \omega_{eg'}$).

In the limit of a weak probe driving field ($\Omega_p \ll \Omega_c, \Delta_c$), the ground state $\ket{g}$ remains an eigenstate with the eigenvalue $\lambda_g = (\Delta_c-\Delta_p$).
The other two eigenstates $\ket{\pm}$ are associated with the two light-shifted resonances close to $\Delta_p=0$ and $\Delta_p=\Delta_c$ as the probe detuning $\Delta_p$ is being varied.
Their corresponding eigenvalues are
$\lambda_{+} = - \delta - i\Gamma_{+}/2$ and $\lambda_{-} = \Delta_c + \delta - i\Gamma_{-}/2$, respectively, with an associated light shift $\delta$ and radiative decays $\Gamma_{\pm}$ ~\cite{spectroStalgies1998}.
For a large detuning $\Delta_c \gg \Omega_c, \Gamma$, these can be obtained through a perturbative expansion of $1/\Delta_c$:
\begin{align}
    \delta &= \frac{\Omega_c^2}{4\Delta_c}\,, \nonumber\\
    \Gamma_{+} &= \Gamma \frac{\Omega_c^2}{4\Delta_c^2}\,, \nonumber \\
    \Gamma_{-} &= \Gamma - \Gamma_+ = \Gamma \big(1-\frac{\Omega_c^2}{4\Delta_c^2}\big)\,.
    \label{eq:eigenstates_simple}
\end{align}
For a larger $\Omega_p$, the probe-induced coupling between $\ket{g}$ and $\ket{e}$ cannot be neglected, and the light shifts and decay rates have been obtained from the steady-state solution for the three-level optical Bloch equation in the vicinity of $\Delta_p=\Delta_c$~\cite{spectroStalgies1998}:
\begin{align}
    \delta &= \frac{\Delta_c}{4\Delta_c^2+\Gamma^2}(\Omega_c^2-\Omega_p^2)\,, \nonumber\\
    \Gamma_{+} &= \frac{\Gamma}{4\Delta_c^2+\Gamma^2}(\Omega_c^2+\Omega_p^2)\,.
    \label{eq:eigenstates}
\end{align}

The narrow resonance associated with $\lambda_{+}$ is shown to exhibit a Fano-shaped profile~\cite{Cohen-Tannoudji_1992} and possess a spectral width $\Gamma_+ \ll \Gamma$ for $\Omega_c,\Omega_p\ll\Delta_c$.
The Fano-type profile manifests in the excitation spectrum of the scattering rate $|T|^2$~\cite{Cohen-Tannoudji_1992}:
\begin{equation}
    |T|^2
    \propto \, \frac{[2\delta/\Gamma_+ + 2(\Delta_p-\Delta_c-\delta)/\Gamma_+]^2}{1+[2(\Delta_p-\Delta_c-\delta)/\Gamma_+]^2}\,,
    \label{eq:fano}
\end{equation}
which matches the form of a typical Fano profile~\cite{Fano1961}.

When including the atomic center-of-mass motion of the atom to the description, the energy change due to recoil from a scattering event should be considered.
For an atom confined in a harmonic potential of frequency
$\omega_\text{trap}$, when the position uncertainty is much smaller than the wavelength of light (Lamb-Dicke limit), the coupling between the motional states and internal energy levels is characterized by the Lamb-Dicke parameter $\eta = |\vec{k}_p-\vec{k}_c| \cos{(\phi)} \,a_0$. Here $\vec{k}_p$ and $\vec{k}_c$ are the wave vectors of the probe and coupling beams, $\phi$ is the angle between $\vec{k}_p-\vec{k}_c$ and the motional axis, and $a_0 = (\hbar/(2m\omega_{\text{trap}}))^{1/2}$ is the position uncertainty of the particle with mass $m$ in the ground state of the harmonic oscillator~\cite{Keitel2000}. 
For an atom initially in the dark internal state and the motional eigenstate $\ket{n}$, the momentum imparted by light when $|\vec{k}_p-\vec{k}_c| \neq 0$ leads to coupling with the bright state $\ket{+}$ of neighboring motional modes $\ket{n \pm 1}$. 
By choosing $\Delta_p = \Delta_c > 0$ and a suitable $\Omega_c$ such that
$\delta = \omega_\text{trap}$, the scattering spectrum can be tailored such that the transition probability of the $\ket{g,n}\rightarrow\ket{+,n-1}$ red sideband transition is greater than the probability of the $\ket{g,n}\rightarrow\ket{+,n+1}$ blue sideband transition.
This results in effective cooling.
A detailed quantitative analysis of the cooling dynamics using a rate equation description is provided in~\cite{Keitel2000,Morigi2003}.

\section{Fano Spectrum}
\begin{figure}    
    \includegraphics[width=0.45\textwidth]{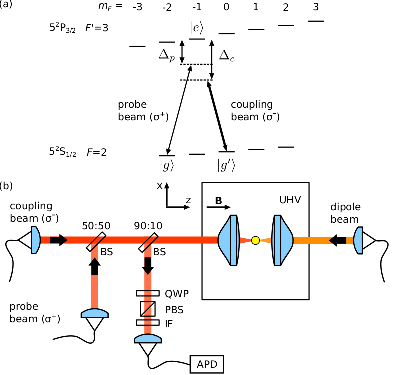}
    \caption{
    (a) Energy levels and transitions in $^{87}$Rb used for observing the Fano scattering profile. 
    (b) Experimental configuration for Fano spectroscopy. 
    The backscattered atomic fluorescence is collected by a high numerical aperture lens and coupled to a single-mode fiber connected to an avalanche photodetector. BS: beamsplitter, QWP: quarter-wave plate, PBS: polarizing beamsplitter, IF: interference filter, APD: avalanche photodetector, UHV: ultra-high vacuum, B : magnetic field.
    \label{fig:setup_spec}}
\end{figure}

To observe the Fano spectrum from a single $^{87}$Rb atom, we consider a $\Lambda$ system formed by the Zeeman sublevels $\ket{g}\equiv\ket{F=2,m_F=-2}$ and $\ket{g'}\equiv
\ket{F=2,m_F=0}$ of the 5$^{2}$S$_{1/2}$ $F$=2 hyperfine ground state and $\ket{e} \equiv
 \ket{F'=3,m_{F'}=-1}$ of the 5$^{2}$P$_{3/2}$ $F'$=3 excited state, subject
 to a pair of laser beams with opposite polarizations (see
 Fig.~\ref{fig:setup_spec}(a)). A stronger left circularly polarized ($\sigma^-$) coupling beam of Rabi frequency $\Omega_c$, couples $\ket{g'}$ to $\ket{e}$ with a detuning $\Delta_c$. 
 A weaker right circularly polarized ($\sigma^+$) probe beam of Rabi frequency $\Omega_p$ and detuning $\Delta_p$ drives the $\ket{g} \leftrightarrow \ket{e}$ transition.

Figure~\ref{fig:setup_spec}(b) shows a schematic of our experimental setup. 
We trap a single $^{87}$Rb atom at the focus of a pair of high numerical-aperture (NA=0.75) aspheric lenses in a far-off-resonant dipole trap (FORT). 
The FORT is formed by a linearly polarized Gaussian laser beam at 851\,nm, tightly focused to a waist of $w_{0}$\,=\,1.1\,$\mu$m. 
The aspheric lenses not only enable tight spatial confinement of the atom in the FORT, but also allow efficient collection of fluorescence from the atom.
Refer to~\cite{Chin2017naturecom} for a complete description of our single atom trap.

For driving the $\Lambda$ system, the coupling and probe beams employed are generated from the same external cavity diode laser.
This ensures a fixed phase relationship between the two driving fields. 
The light from this laser is split into two paths for the coupling and probe beams with the frequency of light independently controlled by an acousto-optic modulator (AOM) in each path. 
The two beams are then overlapped in a beam splitter (BS) and co-propagate to the atom in this part of the experiment. 
The co-propagating configuration minimizes the momentum transfer to the atom ($\Delta \vec{k} = \vec{k}_c - \vec{k}_p = 0$, and equivalently $\eta=0$) via the two-photon process, thereby decoupling the center-of-mass motion from the dynamics and allowing the Fano profile to be resolved.

To prevent probe and coupling beams from entering the detection system, the
atomic fluorescence is collected in the backward direction using a 90:10 BS.
An interference filter (IF) prevents dipole trap radiation from reaching the detectors.
Additionally, we employ a polarization filter consisting of a quarter-wave plate (QWP) and a polarizing beam splitter (PBS) to eliminate scattering from the $\ket{F=2,m_F=-2}\rightarrow\ket{F'=3,m_{F'}=-3}$ cycling transition induced by the strong coupling field.

\begin{figure}
    \includegraphics[width=0.45\textwidth]{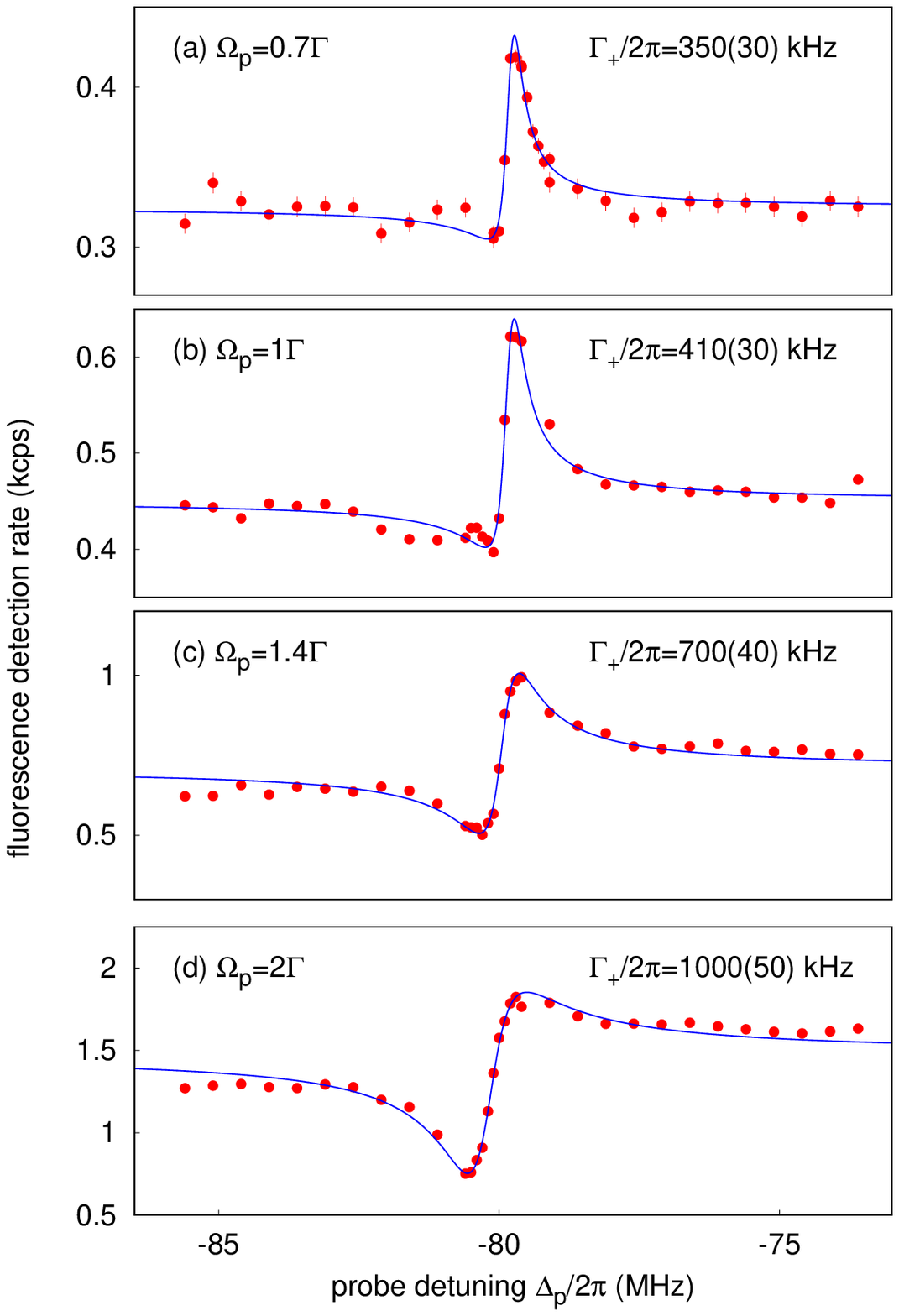}
    \caption{Observation of Fano scattering profiles. 
    Red dots: Single photon scattering detected in APDs from the two-photon process for $\Delta_c/2\pi$\,=\,-\,80\,MHz and $\Omega_c$\,=\,1.4$\Gamma$, projected into the probe polarization. 
    Blue curve: Fits to Fano profiles following Eqn.~\ref{eq:fano}.
    The probe beam power increases from subplot (a) to (d) as indicated by the Rabi frequency values. 
    All plots show a clear suppression in scattering around $\Delta_p/2\pi$\,=\,-\,80\,MHz where the atom is optically pumped to the dark state. 
    Error bars represent one standard deviation due to propagated Poissonian counting statistics.
    \label{fig:scattering}}
\end{figure}

When a single $^{87}$Rb atom is loaded into the FORT, we apply 10\,ms of polarization gradient cooling (PGC) to cool the atom to a temperature of 14.7(2)\,$\mu$K, as measured by the ``release-recapture'' technique~\cite{Tuchendler2008,Chin2017}.
Then, a bias magnetic field of 1.44\,mT is applied along the FORT laser propagation direction to remove the degeneracy of the Zeeman states.
Next, the single atom is illuminated with the pair of strong coupling and weak probe beams for 3\,ms.
During this interval, the atomic fluorescence is detected using an avalanche photodetector (APD).
The measurement is repeated for approximately 3000 runs for various values of $\Omega_p$ as $\Delta_p$ is tuned across a range of $\pm2\pi\times6$\,MHz centered at $\Delta_c$. The coupling beam parameters remain fixed at $\Delta_c = -2\pi\times80$\,MHz and $\Omega_c = 1.4\,\Gamma$.

Figure~\ref{fig:scattering} shows a series of scattering spectra for increasing probe powers.
The detected photoevents shown here also include the APD's dark counts, which contribute to a background of around 300 events per second.
Red points are experimental data and blue lines are fits to Eqn.~\ref{eq:fano}.

In all measurements, an asymmetrical Fano peak is observed with a linewidth smaller than the
natural linewidth ($\Gamma$\,$=$\,$2\pi\times6$\,MHz). The 
Fano linewidths extracted from the fits increase linearly with the probe power
($\Gamma_+/2\pi$ = 350\,(30), 410\,(30), 700\,(40), and 1000\,(50)\,kHz for
saturation parameters of $2\Omega_p^2/\Gamma^2 = 1, 2, 4,$ and 8, respectively).
Compared to the theoretical predictions from Eqn.~\ref{eq:eigenstates}
(yielding $\Gamma_+/2\pi$ $\approx$ 83, 100, 132, and 201\,kHz), the measured
values are larger by a factor of 4.7(6).
This discrepancy could be attributed to the presence of multiple Fano resonances resulting from other Zeeman sublevels. 
Specifically, there is a $\Lambda$ configuration formed by the states $\ket{F=2, m_F=-1}$, $\ket{F'=3, m_{F'}=0}$, and $\ket{F=2, m_{F}=1}$, as well as another $\Lambda$ configuration formed by the states $\ket{F=2, m_F=0}$ ($\ket{g'}$), $\ket{F'=3, m_{F'}=1}$, and $\ket{F=2, m_{F}=2}$.
The coupling strengths are quite different for these $\Lambda$ configurations, which leads to distinct values for shifts and linewidths in the Fano resonances.
Consequently, the scattering profiles for these three sets of $\Lambda$ configurations would overlap and distort the total scattering rate, causing the apparent broadening in the excitation spectrum.

Furthermore, the energy of the dark state indicated by the dip in the scattering spectra, should ideally remain fixed at $\Delta_p = \Delta_c = 2 \pi \times -80$\, MHz, independent of the Rabi frequencies $\Omega_c$ and $\Omega_p$ of the driving fields.
However, we observe that the minimum of the scattering spectra shifts to a larger detuning for increasing $\Omega_p$.
It seems likely that this is because the probe field $\Omega_p$ also drives the transition between the state $\ket{g'} = \ket{F=2, m_F=0}$,  and the excited state $\ket{F'=3, m_{F'}=1}$, which is not taken into account in the three-level model.
This coupling introduces an additional light shift on the $\ket{g'}$ state, leading to a shift in the scattering spectrum for increasing probe field strength.


\section{Cooling of atomic motion}
\begin{figure}    
    \includegraphics[width=0.45\textwidth]{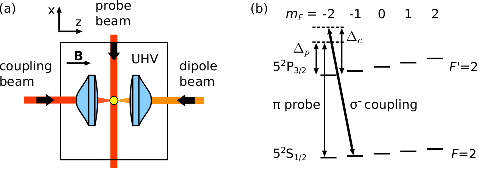}
    \caption{(a) Experimental configuration for the off-resonant EIT cooling process. The probe beam propagates orthogonally to the optical axis to allow for motional coupling.
    (b) Energy levels and transitions in $^{87}$Rb used in the cooling experiment.
    \label{fig:setup_cool}}
\end{figure}

Having developed a better understanding of the absorption profile, we now turn to the cooling of atomic motion. 
In order to utilize the sensitivity of the internal dark state to the spatial gradient of the electric fields, we require a configuration in which the momentum transferred by light to the atom is non-zero ($\Delta \vec{k} = \vec{k}_c - \vec{k}_p \neq 0$). For this, the direction of the probe beam is altered such that it is sent orthogonal to the coupling beam in a top down direction, polarized parallel to the bias magnetic field to excite $\pi$ transitions (see Fig~\ref{fig:setup_cool}).
The $\Lambda$ configuration is now realized with a $\sigma^-$ polarized coupling light connecting $\ket{g'}\equiv\ket{F=2,m_F=-1}$ sublevel of the 5$^{2}$S$_{1/2}$ $F$=2 hyperfine ground state and $\ket{e}\equiv \ket{F'=2,m_{F'}=-2}$ sublevel of the 5$^{2}$P$_{3/2}$ $F'$=2 hyperfine excited state, and a $\pi$ polarized probe light connecting sublevel $\ket{g}\equiv\ket{F=2,m_F=-2}$ of the 5$^{2}$S$_{1/2}$ $F$=2 hyperfine ground state to $\ket{e}$. 
Both coupling and probe are blue-detuned from their respective transitions by $\Delta_c= \Delta_p =2\pi$\,$\times$\,94.5\,MHz\,$\approx$\,16\,$\Gamma$.

Our FORT traps the atom in a 3-D harmonic oscillator with radial ($\omega_{x/y}$) and axial ($\omega_z$) trapping frequencies ($\omega_{x/y}$, $\omega_z$)\,=\,2$\pi\,\times$\,(\,73(2),\,10(1)\,)\,kHz, deduced from a parametric excitation measurement~\cite{Wu2006}. 
Accordingly, the associated Lamb-Dicke parameters ($\eta_x$, $\eta_z$), which quantify the motional coupling, are estimated to be (\,0.23, 0.61\,) for our EIT cooling beam geometry.

\begin{figure}
    \includegraphics[width=0.45\textwidth]{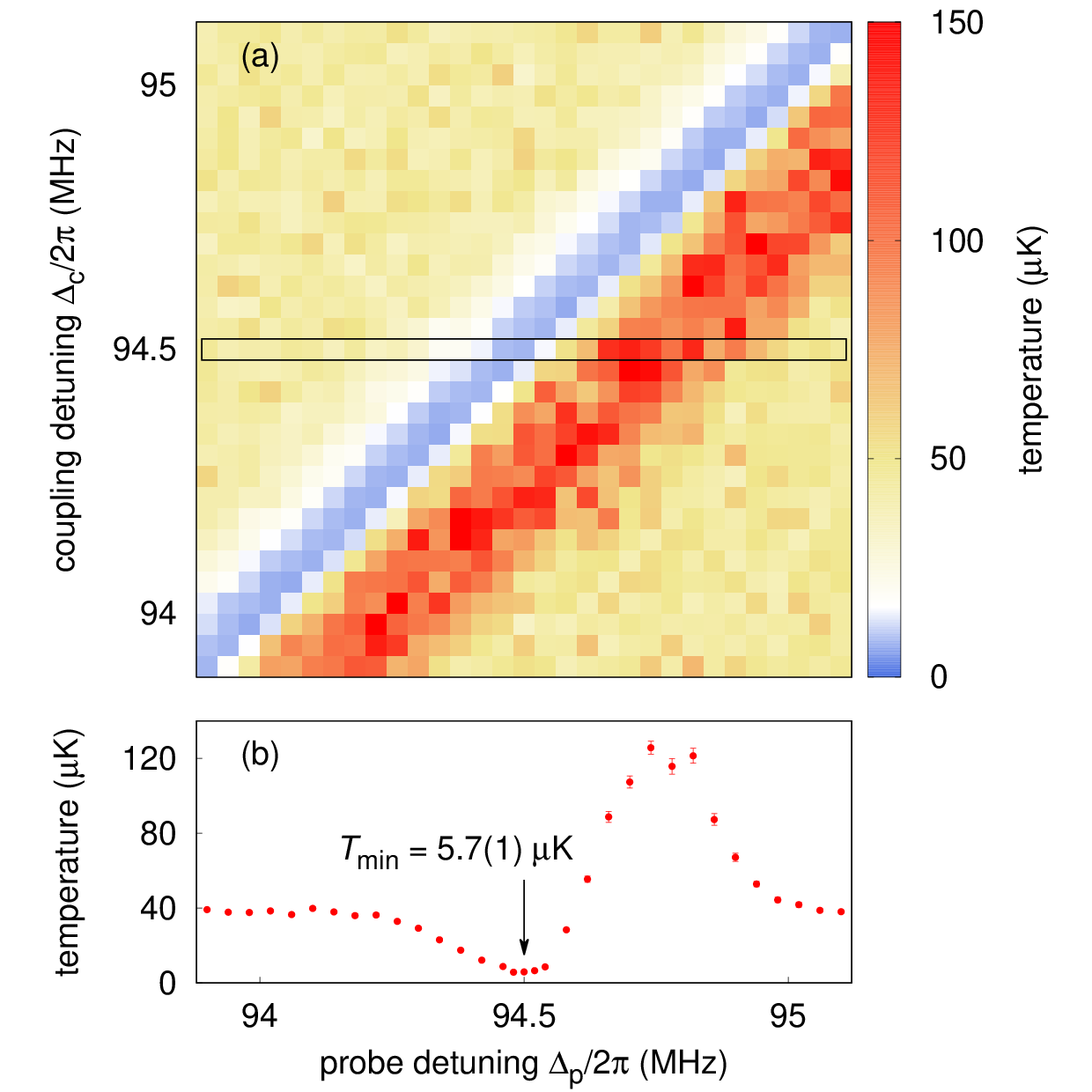}%
    \caption{(a) Atomic temperature for various probe and coupling field
      detunings, inferred from release and recapture measurements after 20\,ms of EIT cooling. The anti-diagonal blue band indicates dark state resonance which has the highest cooling efficiency.
    (b) EIT cooling profile in atomic temperature as a function of probe detuning $\Delta_p$ for a fixed coupling detuning (here  $\Delta_c = 2\pi \times$\,94.5\,MHz as indicated by the boxed region in (a)) also shows an asymmetric Fano feature.
    \label{fig:surv2d}}
\end{figure}

Similar to the experimental sequence described in the previous part, we start with 10\,ms of PGC to cool the atom upon successful loading, followed by a bias magnetic field of 1.44\,mT along the FORT laser propagation direction to remove the degeneracy of the Zeeman states.
We then apply EIT cooling on the $\Lambda$ system by switching on the coupling
beam and probe beam for 20\,ms, a duration chosen to be sufficiently long to ensure that the system reaches a steady state.
During this cooling process, a weak repumper beam resonant to the D1 line at 795\,nm between 5$^{2}$S$_{1/2}\,F=1$ and 5$^{2}$P$_{1/2}\, F'=2$ is also switched on to transfer the atom back into the $F=2$ hyperfine ground state if it spontaneously decays into the $F=1$. 

Following that, we employ a ``release and recapture'' method~\cite{Tuchendler2008,Chin2017} to quantify the temperature of the single atoms.
During this process, the EIT cooling beams are switched off, and the atom is
released from the trap for an interval $\tau_r$ by switching off the FORT beam. 
Subsequently, the FORT is switched on to recapture the atom and we observe atomic fluorescence by switching on the MOT's cooling and repumping beams to check the presence of the single atom.
We repeat each experiment around two hundred times to obtain an estimate of the recapture probability.
We then infer the atomic temperature by comparing the experimentally obtained recapture probability at  $\tau_r$ to Monte Carlo simulations of recapture probabilities for single atoms at various temperatures~\cite{Tuchendler2008}.

In the first part of the thermometric experiment, we investigate the capability of the two-photon process to either cool down or heat up the single atoms.
We apply EIT cooling by varying $\Delta_p$ and $\Delta_c$ over a range of $\pm2\pi\times1$\,MHz while fixing $\Omega_c$ and $\Omega_p$ to $2\pi\times5.2\,$MHz and $2\pi\times2.0$\,MHz, respectively.
We choose $\Omega_c = 2\pi\times5.2\,$MHz because this parameter is expected to give a Fano resonance shift coinciding with the trap frequency ($\delta = \omega_{x/y}$ following Eqn.~\ref{eq:eigenstates_simple}) that leads to optimal cooling.
Here, we fix the release interval to $\tau_r = 30$\,$\mu$s, empirically determined to yield the largest signal contrast for recapturing measurements from which the temperature can be inferred.

The resulting atomic temperature is shown in Fig.~\ref{fig:surv2d}(a).
Cooling and heating effects close to the dressed states for the two-photon process are significantly visible.
We observe an effective cooling in the anti-diagonal stripe where $\Delta_p = \Delta_c$, in agreement with the theoretical prediction.
Heating occurs most dominantly around $\Delta_p = \Delta_c + 2\pi$\,$\times$\,250\,kHz, where the blue sideband transitions have a larger probability.

\begin{figure}
  \includegraphics[width=0.45\textwidth]{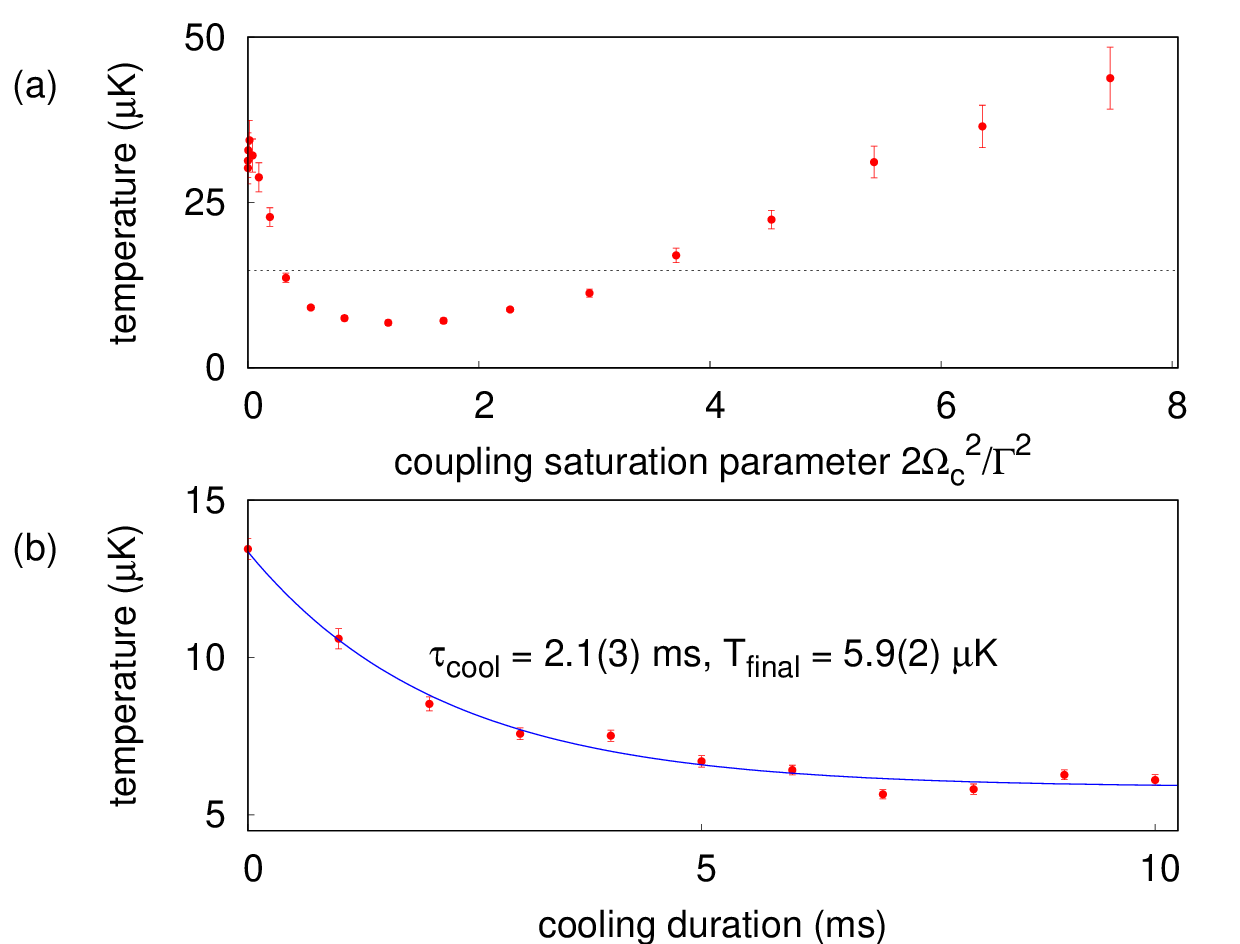}
  \caption{(a) Atomic temperature at $\Delta_p = \Delta_c =
    2\pi\times94.5\,$MHz for varying $\Omega_c$. We observe an effective
    cooling for $s=2\Omega_c^2/\Gamma^2$ between 0.5 and 3, with the optimal
    cooling around $s=1.42(3)$ (cooling duration fixed to 20\,ms).
    The dotted line indicates the initial atomic temperature after PGC of 14.7\,$\mu$K.
    Error bars represent standard error of binomial statistics accumulated from around 200 repeated runs.
    (b) Atomic temperature measured after different cooling durations. A cooling time of 2.1(3)\,ms and final temperature of 5.9(2)\,$\mu$K are extracted from the exponential fit. \label{fig:survpower}}
\end{figure}

In the following parts, we maintain $\Delta_c$ to be fixed at $2\pi\times94.5$\,MHz.
To obtain a more accurate estimation of the atomic temperature, we now deduce a temperature value based on a series of recapturing probabilities for 12 different release intervals, ranging from 1 to 80\,$\mu$s.
We vary the probe detuning $\Delta_p$ around $\Delta_c$, as shown in Fig.~\ref{fig:surv2d}(b).
We observe the typical asymmetric Fano profile also in the temperature of the atoms, with the lowest temperature of 5.7(1)\,$\mu$K measured at $\Delta_p = \Delta_c$. 

We expect optimal cooling to be achieved when the dressed state energy shift $\delta$ caused by the coupling beam is equal to the trap frequency, $\delta=\omega_{x/y}$, as it maximizes the absorption probability on the red sideband transition ~\cite{Morigi2003}. 
To confirm this behavior, we record the atomic temperature using the same
``release and recapture'' scheme for different coupling beam powers, keeping $\Delta_c=\Delta_p=2\pi\times94.5$\,MHz and $\Omega_p=2\pi\times$2.0\,MHz fixed.
The results are shown as a function of the saturation parameter $s=2\Omega_c^2/\Gamma^2$ in Fig.~\ref{fig:survpower}(a).
Cooling is observed for $s$ between $0.5$ and $3$, with the lowest temperature obtained at an optimal cooling parameter of $s=1.42(3)$ (or $\Omega_c=2\pi\times5.06$(5)\,MHz).
This corresponds to a dressed state energy shift of $\delta = \Omega_c^2/(4\Delta_c) \approx 2\pi\times68(1)$\,kHz, as introduced in Eqn.~\ref{eq:eigenstates}, which is comparable with the radial trap frequency $\omega_{x/y}$ in our system.

We then extract the cooling rate by measuring the atomic temperature after a variable time of of EIT cooling, as shown in Fig.~\ref{fig:survpower}(b).
Here, we apply the optimal cooling parameters ($\Delta_c=\Delta_p=2\pi\times94.5$\,MHz, $\Omega_c=2\pi\times5.06$\,MHz and $\Omega_p=2\pi\times$2.0\,MHz) to the pair of coupling and probe beams.
From an exponential fit to the experimental data, we deduce a $1/e$ cooling time constant of 2.1(3)\,ms, and a steady-state temperature of around 5.9(2)\,$\mu$K.

\section{Discussion and Conclusion}
By applying EIT cooling optimized for the radial directions, we have successfully cooled the atom to a temperature of 5.7(1)\,$\mu$K. This is 2.5 times lower than the temperature of 14.7\,$\mu$K typically achieved with conventional PGC. We note that our temperature measurement predominantly reveals the temperature along the radial direction due to the limitation of the ``release and recapture'' technique. Particularly, a Gaussian optical dipole trap typically has a much smaller spatial confinement in the radial direction than in the axial direction. 
Consequently, it is much easier for the atom to escape the trap in the radial direction during the release interval. 

From the measured atomic temperature, we infer a mean phonon number of $\langle n_{x/y} \rangle$ = 1.18(5). 
This is higher than the theoretical value of 0.002 expected for our parameters from the rate equation described in~\cite{Keitel2000}. 
Additionally, we also observe that the measured cooling time constant is about 10 times longer than the expected value of 0.2\,ms estimated from the same theoretical work.
These discrepancies are possibly due to unaccounted heating effects originating from scattering of the strong coupling beam which is red-detuned from the $\ket{F=2,m_F=-2} \leftrightarrow \ket{F'=3,m_{F'}=-3}$ cycling transition. 
In the absence of the EIT cooling, this scattering process alone would impose a lower limit on the energy reached to be in the order of $\sim \hbar \Gamma$ which is  $ \sim 100\,\mu$K in temperature. 
We expect a steady state between these two processes settling at a final temperature approximately two orders of magnitude lower.

In addition, the cooling time would also be limited by the high probability (50\,\%) of an atom in the state $\ket{e}$ of 5$^{2}$P$_{3/2}$ $F'$=2  to decay into the 5$^{2}$S$_{1/2}$ $F$=1 hyperfine level, which is decoupled from the pair of EIT cooling beams. 
Despite the use of a repump light to transfer the atom back to the $F=2$ state,
this process introduces a delay as well as heating. 
In comparison, EIT cooling is 1.9 times slower than the conventional PGC, which has a typical $1/e$ cooling time constant of 1.1(1)\,ms~\cite{Chin2017}.

Although prior work with EIT cooling has demonstrated approximate  ground state occupation, the temperature of  5.7(1)\,$\mu$K achieved here is comparable to the 7\,$\mu$K obtained previously in a standing wave optical trap of~\cite{Meschede2014} and an order of magnitude lower than the temperatures achieved in an ion trap~\cite{Blatt2000}. 
Our demonstration could be extended to lower temperatures further by adding a
second stage of EIT cooling that targets cooling along the axial direction
with $\delta$ matched to the axial trap frequency spacing
$\omega_z$. Exploring strategies to mitigate heating caused by scattering in a multi-level atom could improve the cooling even further.


In conclusion, we have demonstrated electromagnetically induced transparency (EIT) cooling for a single neutral atom confined in a shallow optical dipole trap, and have resolved the  signature Fano profiles in the excitation spectrum
due to a large solid angle for fluorescence collection.
A final temperature of less than 6\,$\mu$K has been reached with EIT cooling, a factor of two below the value obtained by polarization gradient cooling in the same system.

Technologically, the use of magnetic sublevels to realize the $\Lambda$ scheme
is convenient as it requires only a small frequency difference (on the order
of MHz) between the pump and coupling fields, which allows simple frequency shifting from the same laser to provide both components.
This cooling scheme therefore can diversify the spectrum of techniques for manipulation of atomic motion  of ultracold atoms in optical tweezer arrays.

\begin{acknowledgments}
We acknowledge the support of this work by and the National Research
Foundation, Prime Minister's office and A*STAR under project NRF2021-QEP2-01-P01/W21Qpd0101.
\end{acknowledgments}

\bibliographystyle{apsrev4-2}
%

\end{document}